\documentclass[final]{elsart}

\usepackage{amsmath,amssymb,amsfonts}
\usepackage{draftcopy}                  
    \draftcopySetGrey{0.94}             
\usepackage{graphicx}    

\newcommand{\be}{\begin{equation}}
\newcommand{\ee}{\end{equation}}         
\newcommand{\bqa}{\begin{eqnarray}}
\newcommand{\eqa}{\end{eqnarray}}
\newcommand{\bqq}{\begin{eqnarray*}}
\newcommand{\eqq}{\end{eqnarray*}}

\newcommand{\ba}{\begin{array}}
\newcommand{\ea}{\end{array}}

\begin{document}
\begin{frontmatter}



\title{Analytical Solution of the O-X Mode Conversion Problem}
\vspace*{-5mm} 


\author{Francesco Volpe} 
\ead{fvolpe@wisc.edu} 

\address{Dept of Engineering Physics, University of Wisconsin, \\
Madison, WI 53706, U.S.A.}

\begin{abstract}
The excitation of a slow extraordinary wave in a overdense plasma from an
ordinary wave impinging on the critical layer in the plane spanned by 
the density gradient and magnetic field is 
solved analytically by formulating the problem in terms of a parabolic
cylinder equation. A formula for the
angular dependence of the transmission coefficient is derived.
\end{abstract}

\end{frontmatter}


The mode conversion of an ordinary (O) into a slow extraordinary (SX) wave,
first observed in ionospheric plasmas \cite{Budden}, has lately received 
great attention as part of the OXB process \cite{Prein}, where B denotes the
electron Bernstein mode. This has been used with success for heating
\cite{Laqu1}, temperature diagnostic \cite{Laqu2} and current drive
\cite{Laqu3} of the W7-AS stellarator and other hot overdense magnetically 
confined laboratory plasmas \cite{Volpe,LaquPPCF}.

The O-SX mode conversion has been solved in the past under the 
Wentzel-Kramers-Brillouin (WKB) approximation in one dimension (1D)   
\cite{Budden,Prein,Ginzb,Mjol,Kaufm}, in 1D in presence of 
a sheared magnetic field \cite{Cairns}, in two dimensions (2D)  
\cite{Weitzner,Gospod,Popov1} and in 2D in presence of a sheared field 
\cite{Popov2}. Later it was the object of a numerical study \cite{HansFW}. 

The present Letter provides the first {\em analytical}, 
{\em non-WKB}-approximated
solution of the problem. At the cost of a simpler geometry compared 
with Refs.\cite{Cairns,Weitzner,Gospod,Popov1,Popov2}, it leads to a 
relatively simple formula which allows to {\em visualize}  
the spatial dependence (Figs.4 and 6) and time evolution (Fig.5) of the wave as 
it tunnels through the O-SX evanescent layer. This might replace time-consuming 
{\em numerical}  
models in the interpretation of direct, space- and time-resolved measurements 
of the electric and magnetic field in the mode-conversion region \cite{Podoba}.

Consider a plasma magnetized along $z$ and plane-stratified along $x$. 
Suppose the inhomogeneity to be weak enough to use the {\sl local}  
dielectric tensor. 

From the Vlasov-Maxwell system it follows, in the zero temperature limit, that  
\be
\nabla \times (\nabla \times {\bf E}) 
= 
\frac{\omega^2}{c^2} \epsilon {\bf E}, 
\label{fullwave}
\ee 
where $\epsilon$ is the dielectric tensor for a cold 
magnetized plasma \cite{Stix}, 

\be  {\bf \epsilon}=                                  \label{DielCold}
\left( 
\begin{array}{ccc}
1 - \frac{X}{1-Y^2} 
& -i \frac{XY}{1-Y^2}
& 0 \\
i \frac{XY}{1-Y^2}
& 1 - \frac{X}{1-Y^2}
& 0 \\
0 & 0 & 1- X
\end{array}
\right),
\ee

$X=\omega_{pe}^2/\omega^2$ and $Y=\omega_{ce}/\omega$ are the dimensionless 
density and magnetic field, $\omega_{pe}$ and $\omega_{ce}$ the plasma and  
electron cyclotron angular frequency. 
We seek solutions of wave equation \ref{fullwave} in the form:
\be
{\bf E} ( x,y,z,\omega)   =    {\bf E}_0 ( x,\omega) \; e^{ik_z z},      
\label{EqEik}
\ee

where the wavenumber $k_z$ is real-valued. 
Note that this is not an eikonal ansatz and that no 
assumption is made on the dependence of ${\bf E}_0$ on $x$. 
Note also, however, that $k_y$=0, i.e.~that 
the wave is assumed to propagate in the plane spanned by the density 
gradient $\nabla n$ (direction $x$) and by  the magnetic field 
(parallel to $z$). 

With this substitution, Eq.\ref{fullwave} rewrites:
\be
- \frac{\partial ^2}{\partial x^2} 
\left( \begin{array}{c}
0 \\ E_{0y} \\ E_{0z} 
\end{array} \right)
+ i N_z \frac{\omega}{c} \frac{\partial }{\partial x} 
\left( \begin{array}{c}
E_{0z}  \\ 
0   \\ 
E_{0x} 
\end{array} \right)
= 
\frac{\omega^2}{c^2} 
\left( 
\begin{array}{ccc}
1 - \frac{X}{1-Y^2} -N_z^2
& -i \frac{XY}{1-Y^2}
& 0 
\\
i \frac{XY}{1-Y^2}
& 1 - \frac{X}{1-Y^2} -N_z^2
& 0 
\\
0 & 0 & 1- X
\end{array}
\right)
\left( \begin{array}{c}
E_{0x} \\ E_{0y} \\ E_{0z} 
\end{array} \right)  \label{EqafterAlgebra}
\ee

where ${\bf N} =c{\bf k}/\omega$ is the refractive index. 
In the remainder, $x$ will be renormalized to the vacuum wavelength 
by replacing 
\be
x\omega/c \longrightarrow x,    \label{EqNrmlz}
\ee

Circular components $E_{0\pm} = (E_{0x} \pm i E_{0y})/\sqrt{2}$ permit to  
diagonalize the tensor on the right hand side:
\be
\label{polpettone}
- \frac{\partial ^2}{\partial x^2} 
\left( \begin{array}{c}
\frac{E_{0+}-E_{0-}}{2} \\ \frac{-E_{0+}+E_{0-}}{2} \\ E_{0z} 
\end{array} \right)
+ i \frac{N_z}{\sqrt{2}} \frac{\partial }{\partial x} 
\left( \begin{array}{c}
E_{0z}  \\ 
E_{0z}  \\ 
E_{0+}+E_{0-}
\end{array} \right)
= 
\left(
\begin{array}{ccc}
 \frac{X_L-X}{1+Y}      &  0                     &  0     \\
     0                  &  \frac{X_R-X}{1-Y}     &  0     \\
     0                  &  0                     &  X_P-X 
\end{array}
\right)
\left( \begin{array}{c}
E_{0+} \\ E_{0-} \\ E_{0z} 
\end{array} \right)  
\ee

The circularly polarized X-modes and the linearly, parallel to $z$ 
polarized O-mode are indeed the {\em independent} eigensolutions of the 
{\em homogeneous} problem. The derivatives on the left hand side introduce a  
medium {\em inhomogeneity} and, being {\em off-diagonal}, 
they {\em couple} the otherwise independent modes,  
accounting for their {\em conversions}.

The diagonal elements of Eq.\ref{polpettone} have been 
expressed in terms of the cutoff densities

\begin{subequations}
\label{EqRompic0}
\be 
X_R = (1-N_z^2) (1-Y)    
\ee
\be 
X_L =(1-N_z^2) (1+Y)     
\ee
\be 
X_P =1,
\ee
\end{subequations}

These correspond to different parts of the cold dielectric tensor 
($R$, $L$ or $P$, in Stix notation \cite{Stix}) being set to 0. 

$X_R$ is the cut-off for the ``fast'' (in the sense of 
the phase velocity) right-handed (R) X-mode. 
$X_L$ and $X_P$, instead (Eqs.\ref{EqRompic0}b-c), are 
the slow X- and O-mode cutoff densities, $X_{SX}$ and $X_O$ respectively:
\begin{subequations}
\label{EqRompic1}
\be  
X_{SX}=X_L 
\ee
\be
X_O=X_P
\ee
\end{subequations}

if $N_z<N_{z,opt}$. If instead, $N_z>N_{z,opt}$,
\begin{subequations}
\label{EqRompic2}
\be  
X_O=X_L 
\ee
\be
X_{SX}=X_P
\ee
\end{subequations}

Here   
\be
 N_{z,opt}^2 = \frac{Y}{Y+1},     \label{EqNzopt}
\ee

is an optimal value of $N_z^2$ making the O- and slow X-mode degenerate.
$N_{z,opt}$ is the optimal value of $N_z$ yielding complete OX mode conversion 
\cite{Budden,Prein,Ginzb,Mjol}. 
 
From Eqs.\ref{EqRompic0}-\ref{EqNzopt} it can be shown that 
$X_O$ is always the smallest between $X_L$ and $X_P$.

At $X \gg X_R$, where no fast X-mode exists ($E_-$=0) we can restrict our 
attention to:
\begin{subequations}
\label{EqAiry}
\be
  \left( \frac{X_L-X}{1+Y} +\frac{1}{L_n^2}\frac{\partial_X^2}{2} \right) E_+ 
  =
  i\frac{N_z}{L_n} \frac{\partial_X}{\sqrt{2}}E_z
\ee
\be
  \left( X_P -X +\frac{1}{L_n^2}                 \partial_X^2     \right)  E_z 
  =
  i\frac{N_z}{L_n} \frac{\partial_X}{\sqrt{2}}E_+
\ee
\end{subequations}

where the subscript $_0$ has been dropped for brevity and we have changed 
coordinate from $x$ to $X$. $L_n$ denotes the dimensionless local 
density length-scale subject to the same normalization as $x$ 
(Eq.\ref{EqNrmlz}). For example, $L_n$=100 means that it takes 
100/$2\pi$ vacuum wavelengths to go from the $X$=0 to the $X$=1 location. 

Partial derivatives are now taken with respect to $X$, 
which allows not to make specific 
assumptions on the density profile $X(x)$. The magnetic field 
$Y$, on the other hand, will be treated as constant because its variation 
in the density gradient region is negligible in most experiments. 

If the cutoffs are well separated, an O-wave starts fading at the $X$=$X_P$ 
or $X$=$X_L$ location (whichever is smaller) and doesn't reach the other 
cutoff, thus it does not couple with the X-wave.
Under these circumstances, Eqs.\ref{EqAiry} reduce to separate  
Airy equations for $E_+$ and $E_z$, 
the most general solutions of which are linear combinations of $Ai$ and $Bi$ 
Airy functions.  
However, $Bi$ solutions diverge and do not make physical sense here. 
Hence, the factor pertaining to the unphysical solution $Bi$ has to be 0 
\cite{Swan}. As a result, apart from factors,
\begin{subequations}
\label{EqSolAiry}
\be
E_+ = Ai \left[ (X-X_L) \sqrt[3]{\frac{2L_n^2}{Y+1}} \right]
\ee
\be
E_z = Ai \left[ (X-X_P) \sqrt[3]{L_n^2} \right]
\ee
\end{subequations}

\begin{figure}[t]
\includegraphics[scale=1]{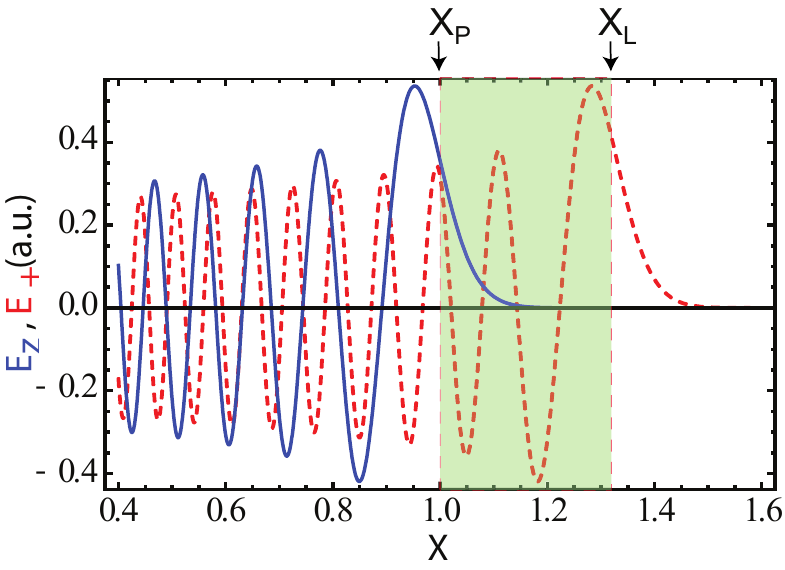}
\caption{\label{FigAiry} 
    O-mode (blue, solid) and X-mode (red, dashed) solutions of 
    Eqs.\ref{EqAiry} when coupling is neglected, for $L_n$=100 and $Y$=0.9, 
    in the vicinity of their respective cutoffs. The colored stripe represents 
    the evanescent region in between.}
\end{figure}

These are the solutions when the O and SX cutoff are well separated. 
As Fig.\ref{FigAiry} illustrates, the O and SX waves evanesce    
exponentially after their cutoffs. Physically, this is because 
the polarization currents (the plasma 
response) grows up to complete cancelation of the displacement current 
$\partial {\bf E} /\partial t$. 
As a result, the O-wave is negligibly small at the other cutoff 
and is not capable of exciting an X-wave. 
This {\em a posteriori} legitimates having neglected the 
coupling terms $iN_z\partial_x/\sqrt{2}$ in Eq.\ref{EqAiry}. 

\begin{figure}[t]
\includegraphics[scale=1]{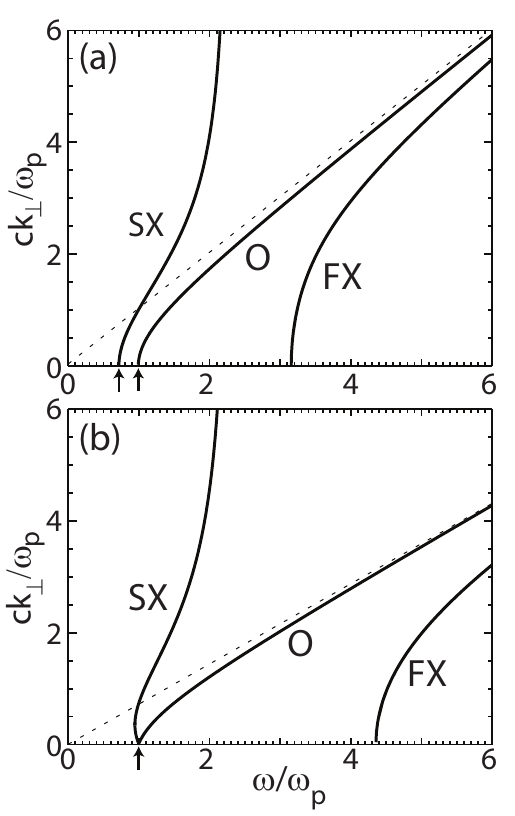}
\caption{\label{FigDisp} 
    Cold plasma dispersion relation of ordinary (O), fast (F) and slow (S) 
    extraordinary (X) waves propagating (a) perpendicularly and (b) with 
    optimal incidence with respect to a magnetic field of strength $Y$=0.9. 
    The dashed line corresponds to propagation in vacuum, for reference. 
    Arrows mark the distinct and degenerate, respectively, O and SX cutoffs. 
}
\end{figure}

Note in Fig.\ref{FigAiry} that the O and SX wave domains of existence are 
$X<X_O$ and $X<X_{SX}$. This is consistent with their frequencies necessarily 
being higher than their respective cutoff frequencies, marked by arrows in 
Fig.\ref{FigDisp}a. For optimal or about-optimal incidence, however, 
the dispersion relation modifies as in Fig.\ref{FigDisp}b, i.e. the frequency 
of the mode-converted SX wave is initially smaller than cutoff, which 
implies access to densities $X>X_{SX}$, up to the turning 
point\cite{Maek,Prein,Weitz} $X=1+Y\frac{1-N_z^2}{2N_z}$.

It is intuitive that when cutoffs are only few decay-lengths far, an O wave 
can excite an SX wave on the other side, and viceversa. 
In this (general) case the coupling terms on the right hand sides of 
Eqs.\ref{EqAiry} have to be retained. 

To eliminate $E_z$, we apply the operator 
$( X_P -X +\partial_X^2 /L_n^2 )$ 
to Eq.\ref{EqAiry}a, the operator 
$  i\frac{N_z}{L_n} \frac{\partial_X}{\sqrt{2}}$
to Eq.\ref{EqAiry}b and take the difference. After some algebra, we obtain: 
$$
\frac{1}{2L_n^4}E_+''''
+\frac{1}{L_n^2}
\left( \frac{X_L-X}{1+Y} +\frac{X_P-X}{2} +\frac{N_z^2}{2} \right) E_+''+
$$
\be
\label{EqODE}
-\frac{2}{L_n^2}\frac{E_+'}{1+Y} 
+\frac{(X_P-X)(X_L-X)}{1+Y} E_+
-i \frac{N_z}{L_n} \frac{E_z}{\sqrt{2}}
=0.
\ee

In the vicinity of cutoffs the partial derivatives $\partial_X$ (proportional 
to $iN_x$) vanish at least linearly if not quadratically with $X$. 
After retaining only the zeros of lower order and taking the real part of 
Eq.\ref{EqODE} while assuming, 
without losing generality, that $E_z$ is real, we obtain:
\be
\label{EqODE2}
N_z^2 \frac{1+Y}{2} E_+'' 
-2E_+' 
+L_n^2 (1-X)(X_L-X) E_+
=0.  
\ee

In Eqs.\ref{EqODE} and \ref{EqODE2}, the symbol 
$'$ denotes derivation with respect to 
$X$. Instead, in terms of spatial derivatives 
$\frac{\partial}{\partial x}$ = $\frac{1}{L_n} \frac{\partial}{\partial X}$, 
\be
\label{EqODE3}
\frac{\partial^2 E_+}{\partial x^2} 
+2 \zeta N_{-,eff} 
\frac{\partial   E_+}{\partial x} 
+ N_{-,eff}^2 E_+
=0
\ee

where we have introduced
\begin{subequations}
\be
\label{EqNeff}
N_{-,eff}^2 = 
\frac{2}{N_z^2} 
\frac{(X_P-X)(X_L-X)}{1+Y}
\ee
\be
\label{Eqzeta}
\zeta = 
-\frac{\sqrt{2}}{L_n N_z} 
[ (1+Y) (X_P-X) (X_L-X) ]^{-1/2}
\ee
\end{subequations}

are respectively an effective squared refractive index and a damping ratio 
for $E_+$.  
Note that both vary with $x$. The spatial non-uniformity of $N_{-,eff}^2$ 
implies that the wavelength changes or, when $N_{-,eff}^2$ becomes 
negative, it implies that the wave 
becomes evanescent. The non-uniformity of $\zeta$, on the other hand, 
determines where $E_+$ decreases, and how strongly. 

Unless the density gradient is extremely steep, thus $L_n$ extremely small, 
the damping ratio $\zeta$ can be ignored 
and Eq.\ref{EqODE3} takes the form 
\be
\label{EqODE4}
\frac{\partial^2 E_+}{\partial x^2} 
+ N_{-,eff}^2 E_+
=0
\ee

From Eq.\ref{EqNeff}, $N_{-,eff}^2$ can be written as 
$aX^2+bX+c$, where $a$, $b$ and $c$ are constant coefficients (remember that 
for simplicity we are assuming uniform $Y$ and slab geometry, thus, conserved 
$N_z$). Therefore, Eq.\ref{EqODE4} can be recognized as a 
parabolic cylinder equation, 
the solutions of which are tabulated \cite{Abram,Zwill1,Zwill2}.

\begin{figure}[t]
\includegraphics[scale=1]{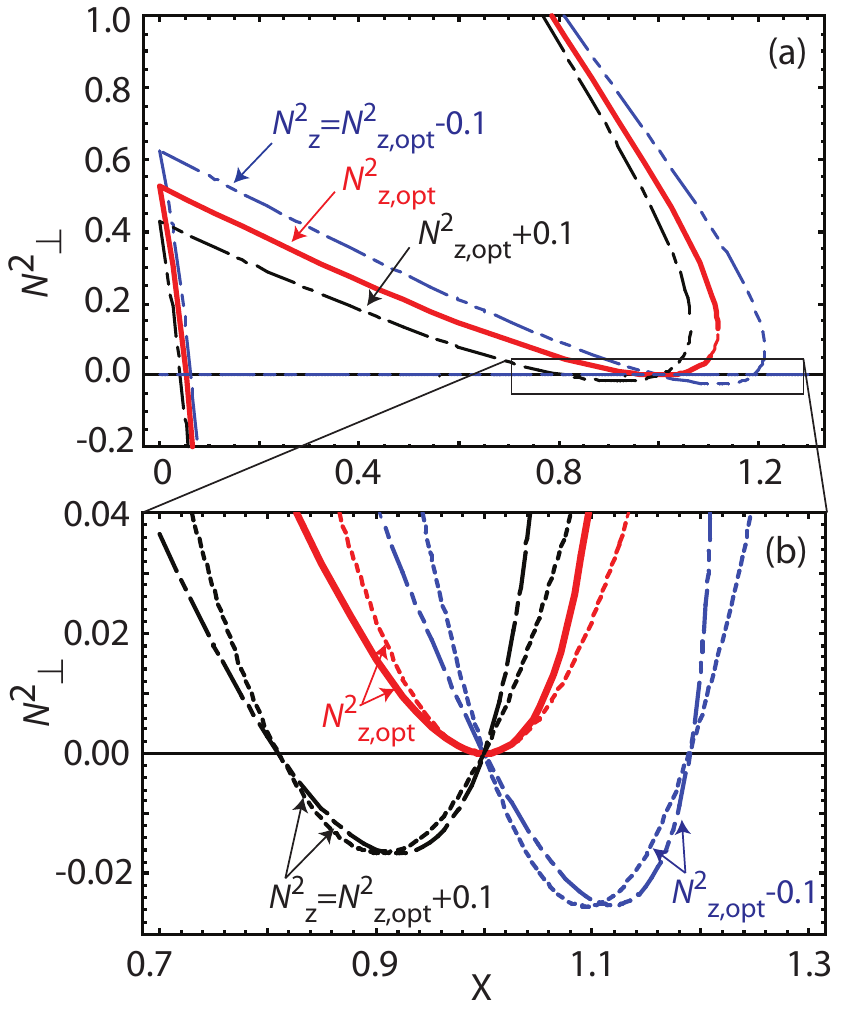}
\caption{\label{FigAppleton} 
    (a) Appleton-Hartree dispersion relation for optimal and slightly 
     non-optimal propagation in $Y$=0.9 and (b) comparison with 
     Eq.\ref{EqNeff} in the vicinity of cutoffs.}
\end{figure}

Eq.\ref{EqNeff} can be thought of as a parabolic ``potential'' in $X$, 
parameterized in $N_z^2$ through $X_L$ and 
locally representing a good approximation (Fig.\ref{FigAppleton}) of 
the Appleton-Hartree dispersion relation\cite{Stix},
\begin{subequations}
\be 
N^2_x +N^2_z=                                         \label{EqAppleton}
1-\frac{2X(1-X)}{2(1-X)-Y^2 N_x^2/N^2 \: \pm \Gamma},
\ee
\be
\Gamma= [ (YN_x/N)^4  \: +4 (1-X)^2 (Y N_z/N)^2]^{1/2}.
\ee
\end{subequations}

For $N_z^2$=$N_{z,opt}^2$ (and only for that value) $X_L$=$X_P$=1. Then 
the parabola in question develops entirely in the positive half-plane 
and the wave is immune from evanescent, $N_{-,eff}^2 <$0 barriers.  

\begin{figure}[t]
\includegraphics[scale=1]{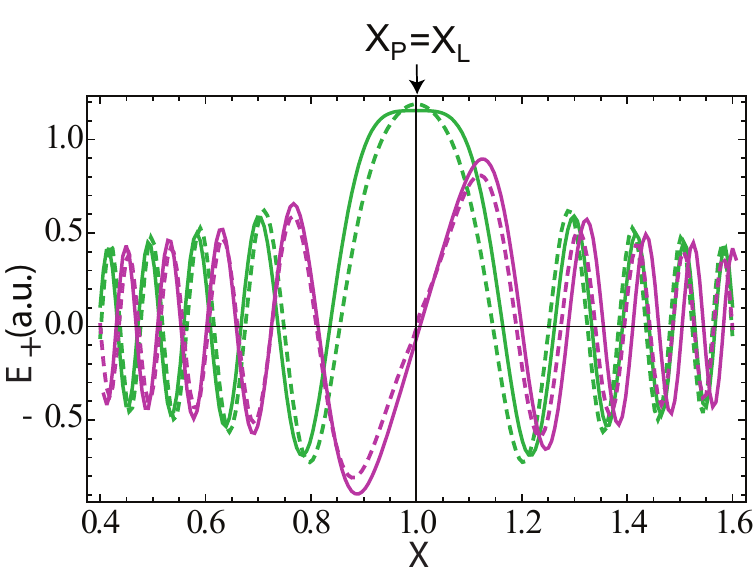}
\caption{\label{FigBess} 
    Odd (purple) and even (green) exact (solid) and approximate (dashed) 
    solutions of the parabolic cylinder Eq.\ref{EqODE4}, 
    for optimal incidence.}
\end{figure}

\begin{figure}[t]
\includegraphics[scale=1]{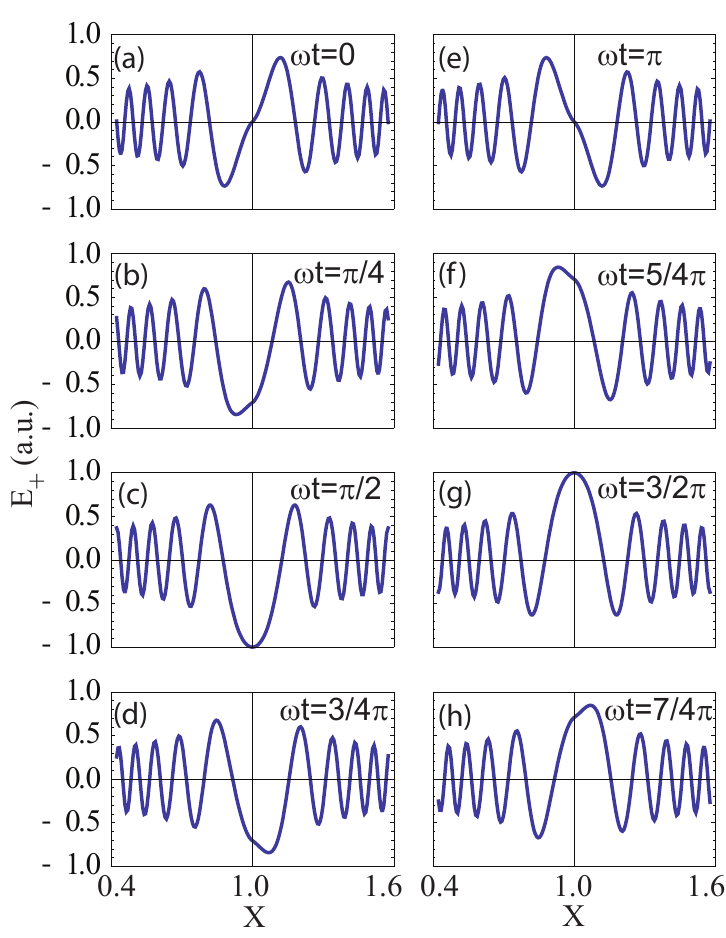}
\caption{\label{FigTimeEvol} 
    Time evolution of a mode-converted O-SX wave crossing the $X$=1 cutoff  
    with optimal incidence.}
\end{figure}

This leads to the simplest solution of Eq.\ref{EqODE4}, 
in terms of parabolic cylinder 
functions of order $-1/2$ \cite{Abram}:
\begin{subequations}
  \label{EqSolA}
  \be C=  
      \left\{
      D_{-\frac{1}{2}} \left[ (-1+i) \sqrt[4]{a}(X-1) \right] +
      D_{-\frac{1}{2}} \left[ ( 1+i) \sqrt[4]{a}(X-1) \right]
      \right\}/2
  \ee 
  \be S=
      \left\{
      D_{-\frac{1}{2}} \left[ (-1+i) \sqrt[4]{a}(X-1) \right] -
      D_{-\frac{1}{2}} \left[ ( 1+i) \sqrt[4]{a}(X-1) \right]
      \right\}/\sqrt{2}
  \ee
\end{subequations}

where 
\be
   a = \frac{2L_n^2}{N_z^2 (1+Y)} 
\ee

reduces to $a=2L_n^2/Y$ in the present, optimal incidence case 
(Eq.\ref{EqNzopt}). The equivalent formulas 
\begin{subequations}
\label{EqBess}
  \be C=  
       a^{1/8} \sqrt{|X-1|} \; J_{-\frac{1}{4}}\left[ \sqrt{a}(X-1)^2/2 \right]
  \ee
  \be S=
       {\rm sign} (X-1) \;
       a^{1/8} \sqrt{|X-1|} \; J_{\frac{1}{4}}\left[ \sqrt{a}(X-1)^2/2 \right]
  \ee
\end{subequations}

involve 1st kind Bessel functions of fractional order. 
These solutions, plotted in Fig.\ref{FigBess} along with their 
approximations 
\begin{subequations}
  \be C \simeq
       \frac{\sqrt[4]{2} \cos \left[ \sqrt{a}(X-1)^2/2 \right]}
            {\left[ \sqrt{a}(X-1)^2 +1\right]^{1/4}}
  \ee
  \be S \simeq
      {\rm sign} (X-1) \;
       \frac{\sin \left[ \sqrt{a}(X-1)^2/2 \right]}
            {a^{1/8} \sqrt{|X-1|}}       
  \ee
\end{subequations}

could be termed ``cosine-like'' and ``sine-like'', for obvious reasons. 
Their linear combination according to inverse prosthaphaeresis-like formulas
\be
\label{EqSolBess}
E_+= C \sin \omega t + S \cos \omega t,
\ee

with weights determined by boundary conditions at a given instant, 
gives snapshots of $E_+$ as a function of $X$ (or equivalently, if the density 
profile is linear, of $x$) at different times $t$.
Here and in Fig.\ref{FigTimeEvol} the phase is chosen to yield a 
sine-like $E_+$ at the time origin $t$=0. 
A tendency to longer wavelength around $X=1$ (a reminescence of reflectometric, 
Airy function behaviour) can be recognized in these graphs.  
Actually $\lambda \rightarrow \infty$ in that point (and only in that point). 
This is clear because that point is a cutoff, defined by $k_x \rightarrow 0$. 

For non-optimal incidence $X_L \ne X_P$ and it is convenient to introduce the 
intermediate density and the semi-difference: 
\begin{subequations}
  \be
     X_m=(X_L+X_P)/2 =1+\delta X
  \ee
  \be
     \delta X = (X_L-X_P)/2 = -(1+Y) \; \Delta N_z^2 /2
  \ee
\end{subequations}

where $\delta X$ is proportional to the thickness of the barrier 
and the systematic error in the squared refractive index, 
$\Delta N_z^2=N_z^2-N_{z,opt}^2$, is related to the systematic 
error in the launching angle. 

\begin{figure}[t]
\includegraphics[scale=1]{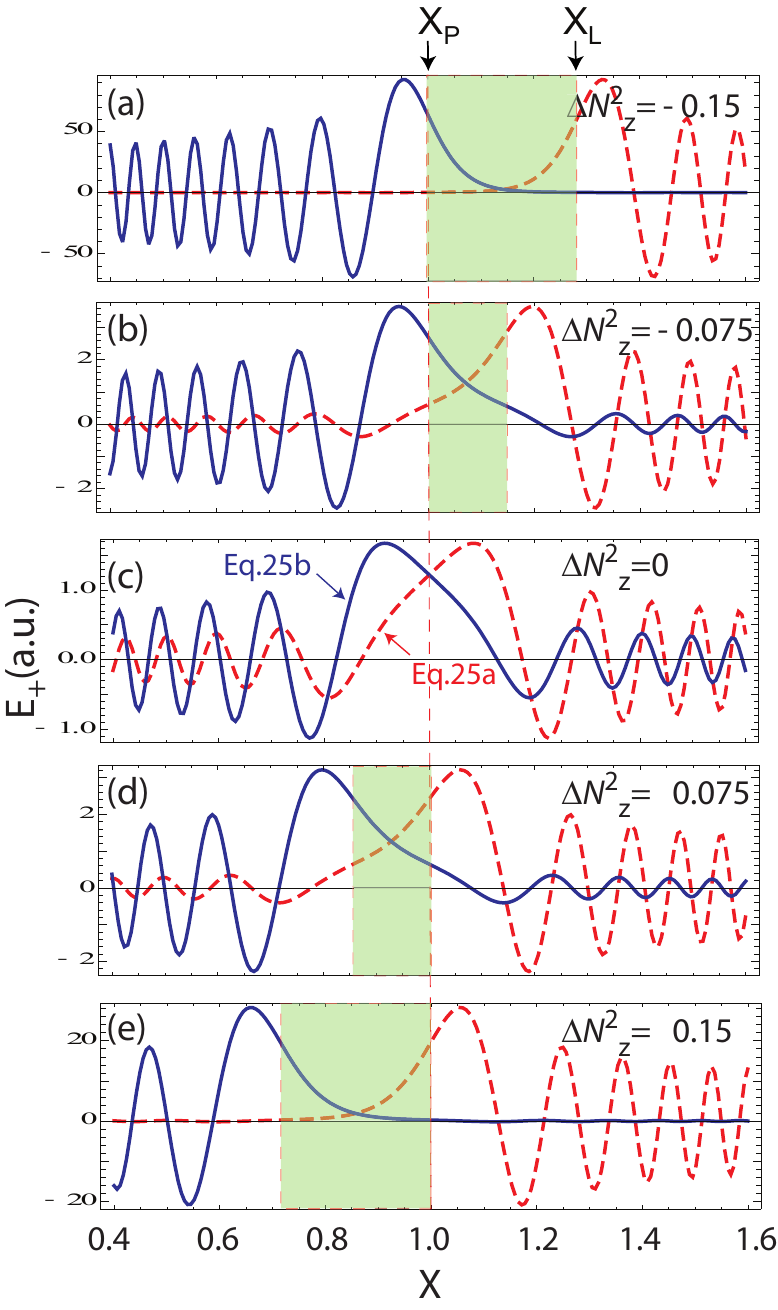}  
\caption{\label{FigVarBarr} 
    Parabolic cylinder solutions, Eq.\ref{EqSolFinal}, for values 
    of $\Delta N_z^2$=$N_z^2-N_{z,opt}^2$ corresponding to 
    (a) far too perpendicular, (b) 
    slightly too perpendicular, (c) optimal, (d) slightly too shallow and 
    (e) too shallow incidence on the evanescent barrier (green).}
\end{figure}

For finite $\delta X$, the parabolic functions involved in Eqs.\ref{EqSolA} 
generalize as follows \cite{Abram,Zwill1,Zwill2}:
\begin{subequations}
  \label{EqSolFinal}
  \be
      E_+ = D_{-\frac{1}{2}-\frac{i}{2}\sqrt{a}(\delta X)^2} 
      \left[ (-1+i) \sqrt[4]{a}(X-1-\delta X) \right]
  \ee
  \be
      E_+ = D_{-\frac{1}{2}+\frac{i}{2}\sqrt{a}(\delta X)^2} 
      \left[ ( 1+i) \sqrt[4]{a}(X-1-\delta X) \right]
  \ee
\end{subequations}

and are plotted in Fig.\ref{FigVarBarr}. 
When incidence is far from optimal, 
Airy functions are re-obtained (Fig.\ref{FigVarBarr}a and e). 
Improved launching conditions thin the evanescent barrier. 
The coupling terms in Eqs.\ref{EqAiry} start playing a role and each Airy 
function acquires an oscillatory nature beyond its respective cutoff as 
a result of the interaction with the other (Fig.\ref{FigVarBarr}b and d).
Optimizing $N_z$ suppresses the barrier, maximizes the mutual influence 
between the O and SX solution and, thus, transmission (Fig.\ref{FigVarBarr}c). 

Transmissivity through the barrier can be defined as the ratio between the 
amplitudes at the barrier edges, i.e.~at cutoffs, squared:
\be
\label{EqTransm}
T=\frac{D^2_{-\frac{1}{2}-\frac{i}{2}\sqrt{a}(\delta X)^2} 
      \left[ ( 1-i) \sqrt[4]{a} \: \delta X \right]}
{D^2_{-\frac{1}{2}-\frac{i}{2}\sqrt{a}(\delta X)^2} 
      \left[ (-1+i) \sqrt[4]{a} \: \delta X \right]}
\ee

\begin{figure}[t]
\includegraphics[scale=1]{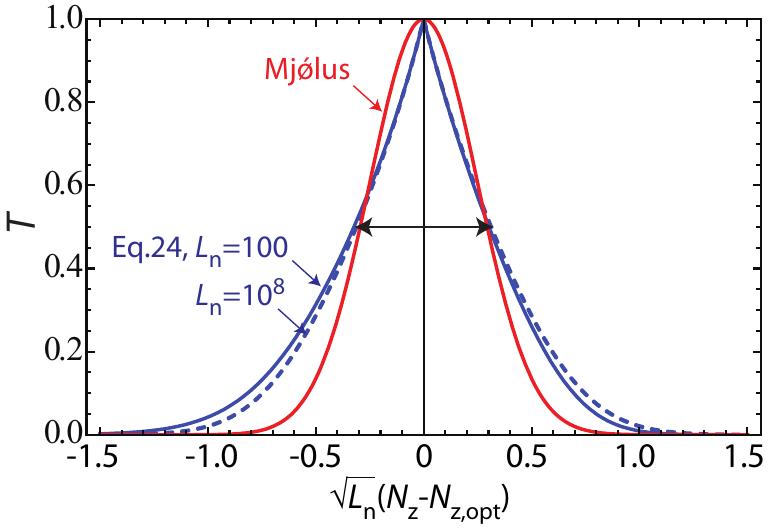}
\caption{\label{FigTransm} 
    Transmissivity inferred from electric field values at the barrier edges in 
    Fig.\ref{FigVarBarr} (blue) and comparison 
    with Mj\o lus' transmissivity (red). The latter preserves its shape, the 
    former nearly preserves it even when the density lengthscale 
    varies by several orders of magnitude.}
\end{figure}

This is plotted in Fig.\ref{FigTransm} and is in reasonable agreement with 
Mj\o lus formula \cite{Mjol}, 
\be
T=\exp 
\left[
  -\pi L_n \sqrt{2Y} 
  (1+Y)(N_{z,opt}-N_z)^2 
\right],
\label{EqMjol}
\ee

here rewritten for the present geometry ($N_y$=0) and normalizations (with 
$L_n$ corresponding to $k_0L_n$ of Mj\o lus). In turn, this was 
validated against full wave calculations \cite{HansFW,Koehn} and agreed  
within error bars with experiments \cite{Laqu1,Laqu2,Shev,Volpe}.
In particular the two curves in Fig.\ref{FigTransm} exhibit the same width 
at half maximum. They also practically exhibit the same dependence on 
$L_n$ (note the renormalized quantity on the horizontal axis), with the 
angular tolerance augmenting if the density gradient steepens. 

In conclusion in the present Letter the problem of the mode conversion of an 
ordinary into an extraordinary mode has been formulated in terms of the 
parabolic cylinder Eq.\ref{EqODE4}. Their solutions for 
optimal (Eqs.\ref{EqBess}) and non-optimal (Eqs.\ref{EqSolFinal}) 
incidence on the cutoff layer allow to visualize the wave behavior in the 
vicinity of and across the degenerate cutoff 
(Fig.\ref{FigBess} and \ref{FigTimeEvol}) and evanescent barrier 
(Fig.\ref{FigVarBarr}), respectively, and with realistic amplitudes, 
in agreement with a well-validated expression for transmissivity 
(Fig.\ref{FigTransm}).

\newpage 

\bibliographystyle{elsart-num}
\bibliography{An2}

\newpage

\end{document}